\documentclass{aa}  
\usepackage[varg]{txfonts}
\usepackage[english]{babel}
\usepackage{amsmath}
\usepackage{graphicx}

\usepackage{natbib}

\include{definitions}

      \def\new#1 {{\bf #1 }}
      \def\cut#1 {\sout{#1} }

\def\kms {$\mathrm{km\,s^{-1}}$} % km s^-1
\def\degr {\hbox{$^\circ$}}
\def\percc {$\mathrm{cm^{-3}}$} %cm^-3
\def\cmsq  {$\hbox{{\rm cm}}^{-2}$}    %cm-2
\def\cmsqg  {$\hbox{{\rm cm}}^{2}$/g}    %cm-2
\def\Lsol {$\hbox{L}_\odot$}
\def\Msol {$\hbox{M}_\odot$}
\def\Blos {${B}_{\rm los}$}
\def\Bpos {${B}_{\rm pos}$}

\def\HII {H{\sc ii}} % H II
\def\AMM {$\mathrm{NH_3}$} %NH3
\def\DAMM {$\mathrm{NH_2D}$} %NH2D
\def\muG {$~\rm{}\mu{}G$} %microG
\def\vlsr {$\mathrm{V_{lsr}}$} %V_LSR  
\def\simgreat{\mathbin{\lower 3pt\hbox
     {$\rlap{\raise 5pt\hbox{$\char'076$}}\mathchar"7218$}}}
\def\simless{\mathbin{\lower 3pt\hbox
     {$\rlap{\raise 5pt\hbox{$\char'074$}}\mathchar"7218$}}}

\newcommand\blfootnote[1]{%
  \begingroup
  \renewcommand\thefootnote{}\footnote{\hspace{-1em}#1}%
  \addtocounter{footnote}{-1}%
  \endgroup
}
\renewcommand{\d}{\mathrm{d}}

\begin{document}

\title{CN Zeeman and dust polarization in a high-mass cold clump}

 %  \subtitle{I. Overviewing the $\kappa$-mechanism}

\author{T. Pillai
  \and
  J. Kauffmann %\fnmsep \thanks{Just to show the usage  of the elements in the author field}
  \and
  H. Wiesemeyer %\fnmsep \thanks{Just to show the usage  of the elements in the author field}
  \and
  K.M.Menten %\fnmsep \thanks{Just to show the usage  of the elements in the author field}
}

\institute{Max-Planck-Institut f\"ur Radioastronomie,
  Auf dem H\"ugel, 53121, Bonn, Germany\\
  \email{tpillai.astro@gmail.com}}

\date{Received XXX; accepted XXX}

\abstract{We report on the young massive clump (G35.20w) in  W48 that previous
  molecular line and dust observations have revealed to be in the very early stages of star
  formation. Based on virial analysis, we find that
  a strong field of ~1640\muG\ is required to keep the clump  in
  pressure equilibrium.  We performed a deep Zeeman effect measurement of the
  113 GHz CN (1-0) line towards this clump with the IRAM 30 m
  telescope. We combine simultaneous fitting of all CN hyperfines with
  Monte Carlo simulations for a large range in realization of the magnetic field  to obtain a
  constraint on the line-of-sight field strength of
  $-687 \pm 420$\muG.  We also analyze archival dust polarization
  observations towards G35.20w.  A strong magnetic field is implied by
  the remarkably ordered field orientation that is perpendicular to
  the longest axis of the clump. Based on this, we also estimate the
  plane-of-sky component of the magnetic field to be $\sim 740$\muG.
  This allows for a unique comparison of the two orthogonal
  measurements of magnetic field strength of the same region and at
  similar spatial scales. The expected total field strength shows no
  significant conflict between the observed field and that required for
  pressure equilibrium.  By producing a probability
  distribution for a large range in field geometries, we show that
  plane-of-sky projections are much closer to the true field strengths
  than line-of-sight projections. This can present a significant
  challenge for
  Zeeman measurements of magnetized structures, even with ALMA. We
  also show that CN molecule does not suffer from depletion on the observed
  scales  in the predominantly cold and highly deuterated core in
  an early stage of high-mass star
  formation and is thus a good tracer of the dense gas.}


\maketitle
\blfootnote{Based on observations carried out with the IRAM 30m
    Telescope. IRAM is supported by INSU/CNRS (France), MPG (Germany)
    and IGN (Spain).}
%
%________________________________________________________________
\section{Introduction}

   High-mass stars ($\ge 8$\,\Msol) exhaust lighter
elements and initiate carbon fusion.  The mass functions of stellar
populations in widely different environments reveals that such stars are
far fewer in number than low-mass stars \citep{zinnecker2007:araa}. Yet, they manage to dominate
the ISM energetics and influence the galaxy as a whole throughout
their evolution.  Therefore it is important to understand how these
stars form; indeed, there have been tremendous strides in the field
of high-mass star formation in the last decade. Multi-wavelength
surveys of the whole or large parts of the Galactic plane, followed up by targeted observations in the last
decade have revolutionized our understanding of their formation (see
\citealt{tan2014}). Such studies have revealed that typical high-mass
protostellar cores are highly Jeans unstable. Therefore,
theoretically turbulence is expected to have a significant influence
in forming high-mass stars \citep{mckee02:100000yrs}. Observations of high-mass star-forming
regions show supersonic line widths in molecular gas tracers
supporting this scenario. Magnetic fields may be equally important in such
regions. However, direct measurements of line-of-sight (los) magnetic fields appear
to show that such regions are magnetically super-critical and thus, do
not favor this scenario (see \citealt{crutcher2012}).  However these
measurements generally have targeted clumps that contain already formed stars whose 
feedback might have altered the initial properties.

Largely unexplored however, are the the initial conditions of high-mass
star formation. Pristine prestellar clumps may be found  in the
vicinity of very young high-mass protostars (with little feedback
effects), which should represent such initial conditions with material in a
cold (10 -- 20 K) and quiescent phase. There is therefore considerable
current interest in the study of such high-mass clumps.  Recently, a
few studies have been conducted at high resolution, involving a
one-to-one comparison of mass derived from observations of thermal dust continuum and virial mass,
to understand whether such clumps are in virial equilibrium \citep{pillai2011a,kauffmann2013b,tan2013:hmsc}. Such studies find that the most massive clumps are
unstable to collapse, unless an enhanced background magnetic field ($\ge
500$\muG) exists. Given this, a determination of the magnetic
field strength in these sources is highly desirable.

In contrast with parameters like density, temperature, velocity field, and molecular abundances, it has been notoriously
difficult to measure $B$-fields in any regime of the interstellar medium (ISM)
from diffuse clouds to dense star-forming cores (see \citealt{crutcher2012,li2014:ppvi_bfield}).
The two main methods to determine magnetic field strength are dust
polarization and  Zeeman measurements. Dust polarization measurements
along with velocity dispersions can be used to determine the $B$-field by
gauging the dispersion in the polarization vectors. A more direct
method of measuring the field strength $\vec{B}$ is via the Zeeman effect, which causes a frequency
shift between the left-hand and right-hand circularly polarized components of a spectral line with a suitable electronic structure.
CN is one such molecular gas tracer, whose  ground-state
hyperfine structure (hfs) transitions near 113~GHz can be detected in clouds with  densities $\ge$ a few
$10^4$ cm$^{-3}$.  This density regime is the most crucial  for the onset of star formation. The Zeeman signal however is exceedingly weak,  making its detection extremely difficult, even for relatively strong lines. 
Therefore, all CN detections to date have been made in very active
high-mass star-forming regions with evolved high-mass protostars that
have bright lines, and  strong fields (\citealt{crutcher1999},
\citealt{falgarone2008}). 

Here, we present magnetic field observations of a very young high-mass clump. The
observations include Zeeman measurements from the IRAM 30 m telescope and
archival dust polarization observations. The target selection and
observations are detailed in section~\ref{sec:targets} and
\ref{sec:obs}, respectively. Zeeman data reduction and analysis
followed by results from the polarization data are reported in
Section~\ref{sec:res}.

%__________________________________________________________________
\section{The target: G35.20w \label{sec:targets}}
\begin{figure}
   \centering
\includegraphics[width=\linewidth]{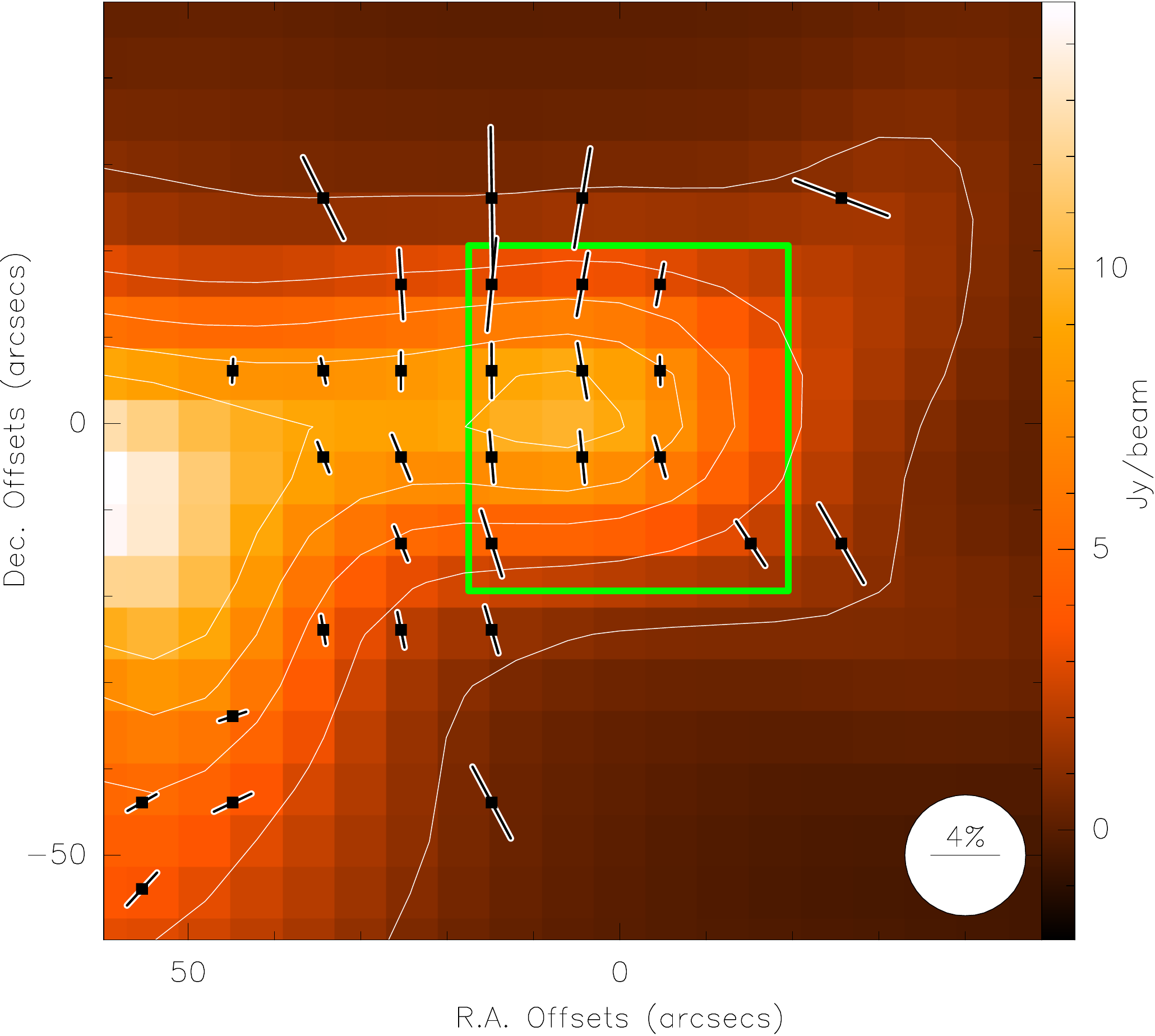}
\caption{G35.20w SCUBA archival $850~\rm{}\mu{}m$ polarization vectors
  rotated by $90^{\circ}$ to show the magnetic field orientation, overlaid
  on SCUBA $850~\rm{}\mu{}m$ dust continuum archival data (color scale).  The
  position offsets are w.r.t the coordinates given in
  Table~\ref{tab:datsum}. The box shows the region around our
  IRAM 30 m CN observations to which we confine our
  analysis. Polarization data is only shown where the polarization
  level is $\ge{}3\sigma$, corresponding to an uncertainty in
  position angle $\le 9.5^{\circ}$ . The length of the \Bpos\ vector represents the percentage
of polarization. \label{fig:polmap}}
   \end{figure}

Our Zeeman target \object{G35.20w} (see Figure~\ref{fig:polmap}) lies about 1\arcmin\, west of the W48
\HII\ region (G35.20$-$1.74).  We adopt a distance of 3.27\,kpc based
on trigonometric maser parallax measurements toward G35.20-1.74
\citep{zhang09_g35}.  The
dust polarization data has been published by \citet{curran2004} where
our target is identified as W48W. Our interferometer 3\,mm dust continuum and
\DAMM\ observations have revealed that the clump  consists of several
massive ($120$\,\Msol\ on average)
and cold cores. \citet{pillai2011a}  find these cores to be highly supercritical. The region is infrared quiet except for the
most massive core, which hosts an embedded protostar driving a massive outflow
\citep{pillai2011a} (hereafter P11). The derived core masses exceed the mass-size threshold for high-mass star formation
\citep{kauffmann2010c, kauffmann2010a:mass_size1} by at least a factor
of 10.  The clump is characterized by high deuteration and low gas temperatures $\sim
20$\, K derived from \AMM\ measurements of P11. The extreme
youth of the clump is  further confirmed
by \cite{rygl2014} who show that the cold cores in this region are dense
structures with the lowest dust temperature ($< 20$\,K) of the whole
W48A environment; these authors base their findings on Herschel dust temperature and column
density measurements.  Rygl et al. also derive a bolometric luminosity of
4000\,\Lsol\ (Clump H3 in their terminology) with an estimated age of
$< 0.2$\,Myrs. Thus, all observations point to G35.20w being a
massive star-forming clump in its youngest stage of evolution.

SCUBA dust continuum data centered on G35.20w is represented in Figure~\ref{fig:polmap}. 
A summary of the  data discussed in this paper is given in Table~\ref{tab:datsum}.

%                                                One column figure
%----------------------------------------------------------- a

%
%______________________________________________________________

\section{Observations  \label{sec:obs}}

\subsection{IRAM 30 m Zeeman observations }

We started with pointed observations toward G35.20w in the CN $N=1-0$ line at 113\,GHz with the IRAM
30 m telescope. Then, we performed a five-point map around the peak. The beam at this
frequency is 23.5\,\arcsec (FWHM). In Table~\ref{tab:freqs}, we give the frequencies of the relevant hfs
transitions that have significant Zeeman splitting ( N=1--0 transition).
After reducing this data, we concluded that the submm clump is
characterized well by the
CN emission  (see section \ref{sec:cn}) and that the
optical depth of CN hyperfine structure (hfs) components is moderate enough to allow for a reliable Zeeman
fitting.  We carried out the Zeeman observations with the XPOL setup
\citep{thum2008}  over
several days in April to May 2010 in good 3\,mm weather conditions
(2-3\,mm water vapor). The final rms on the
peak position is 6\,mK.
 The data were analyzed in CLASS\footnote{http://www.iram.fr/IRAMFR/GILDAS/}. 

\begin{table*}
\caption{Data summary}
\begin{center}
\begin{tabular}{cccccc}
\hline 
Source & $\alpha$ (J2000) & $\delta$ (J2000) & \vlsr\ (\kms) &Dust Polarization  &  Zeeman \tabularnewline
\hline 
G35.20w & 19:01:42.11 & +01:13:33.4 & 42.4 & SCUPOL (JCMT)$^a$ & XPOL
                                                                 (IRAM
                                                                 30 m)\tabularnewline
\hline 
\end{tabular} \\
$^a$: \citep{curran2004, matthews2009}
\end{center}
\label{tab:datsum}
\end{table*}

\begin{table*}
\centering
\caption{CN $N=1-0$ transitions with significant Zeeman splitting}
\begin{center}
\begin{tabular}{rcccc}
\hline
Line &  Transition                         & Frequency  & Zeeman
                                                          Splitting &
                                                                      Intensity$^1$\\
&N,J,F $\rightarrow$ N$'$,J$'$,F$'$ &  GHz       &  Hz/\muG\   \\
\hline
 
      1    & 1,1/2,1/2  $\rightarrow$  0,1/2,3/2   & 113.14416 & 2.18
                                                                    & 8\\
      2    & 1,1/2,3/2  $\rightarrow$  0,1/2,1/2   & 113.17049 & -0.31
                                                                    &
                                                                      8 \\
      3    & 1,1/2,3/2  $\rightarrow$  0,1/2,3/2   & 113.19128 & 0.62
                                                                    & 10 \\
      4    & 1,3/2,3/2  $\rightarrow$  0,1/2,1/2   & 113.48812 & 2.18
  & 10\\
      5    & 1,3/2,5/2  $\rightarrow$  0,1/2,3/2   & 113.49097 & 0.56
                                                                    &
                                                                      27 \\
      6    & 1,3/2,1/2  $\rightarrow$  0,1/2,1/2   & 113.49964 & 0.62
                                                                    & 8 \\
      7    & 1,3/2,3/2  $\rightarrow$  0,1/2,3/2   & 113.50891 & 1.62
                                                                    &
                                                                      8 \\
\hline
\end{tabular}
\end{center}
$^1$: Intensity is the relative intensity of the respective hyperfine transitions. 
\label{tab:freqs}
\end{table*}

\subsection{SCUBA archival dust polarization observations}
\citet{thompson2005:dmuconf}  reported the SCUBA observations that resulted in our dust continuum data.
We also analyze archival calibrated polarization data obtained with
the SCUBA instrument at the JCMT. Specifically, we use data from the SCUPOL catalog, which is a compilation of
850$\mu$m polarization observations at JCMT \citep{matthews2009}. We
refer to Matthews et al. for a description of the observations and data
reduction. Data for G35.20w has already been published
by \citet{curran2004}. They derive a magnetic field strength and assess the
mass-to-flux ratio using the Chandrasekhar-Fermi method, however with a
highly uncertain mass estimate and an assumed velocity dispersion. With
temperature, density and velocity dispersion measurements from our
previous work at high as well as low resolution (P11), we are now able to obtain as accurate an estimate of
the field strength as possible.

\section{Results \label{sec:res}}

 \subsection{Dense clump properties \label{sec:crit}}

Here, we revisit the P11 results for G35.20w. We recalculate the parameters
of the clump  now extracted only toward the region around our IRAM 30 m CN observations. We
have thus chosen to exclude the secondary peaks detected in our larger 
dust continuum image. The region thus confined is shown as a green box in
Figure\,\ref{fig:polmap}.  We extract the total flux within that region
from SCUBA dust continuum observations at 850$\mu$m to determine the total mass. The
virial mass is characterized by the line width (2.1\,\kms) associated with
the systemic velocity component (42.4\,\kms) over the same region using
our CN observations following the formulation in P11. The effective radius estimated
from the source area A (green box in Figure\,\ref{fig:polmap}) calculated as $\sqrt{(A/\pi)}$ is $\sim 0.35$\,pc.  We
derive a virial mass of 347\,\Msol. The average gas temperature from previous \AMM\
observations is 20\,K (P11). Assuming the same temperature for dust as
confirmed by \cite{rygl2014} and a
dust opacity of 0.02\,\cmsqg\ (\citealt{ossenkopf1994:opacities}; thin
ice mantles at n(H)$=10^6$\,\percc),
we derive a mass of 1650\,\Msol\ following \citet{kauffmann2008}.  For reference, the parameters
extracted for the region are also listed in
Table\,\ref{tab:mass}. 

\subsection{CN Zeeman results \label{sec:cn}}

\begin{figure*}
\begin{center}
\includegraphics[angle=0,scale=0.40]{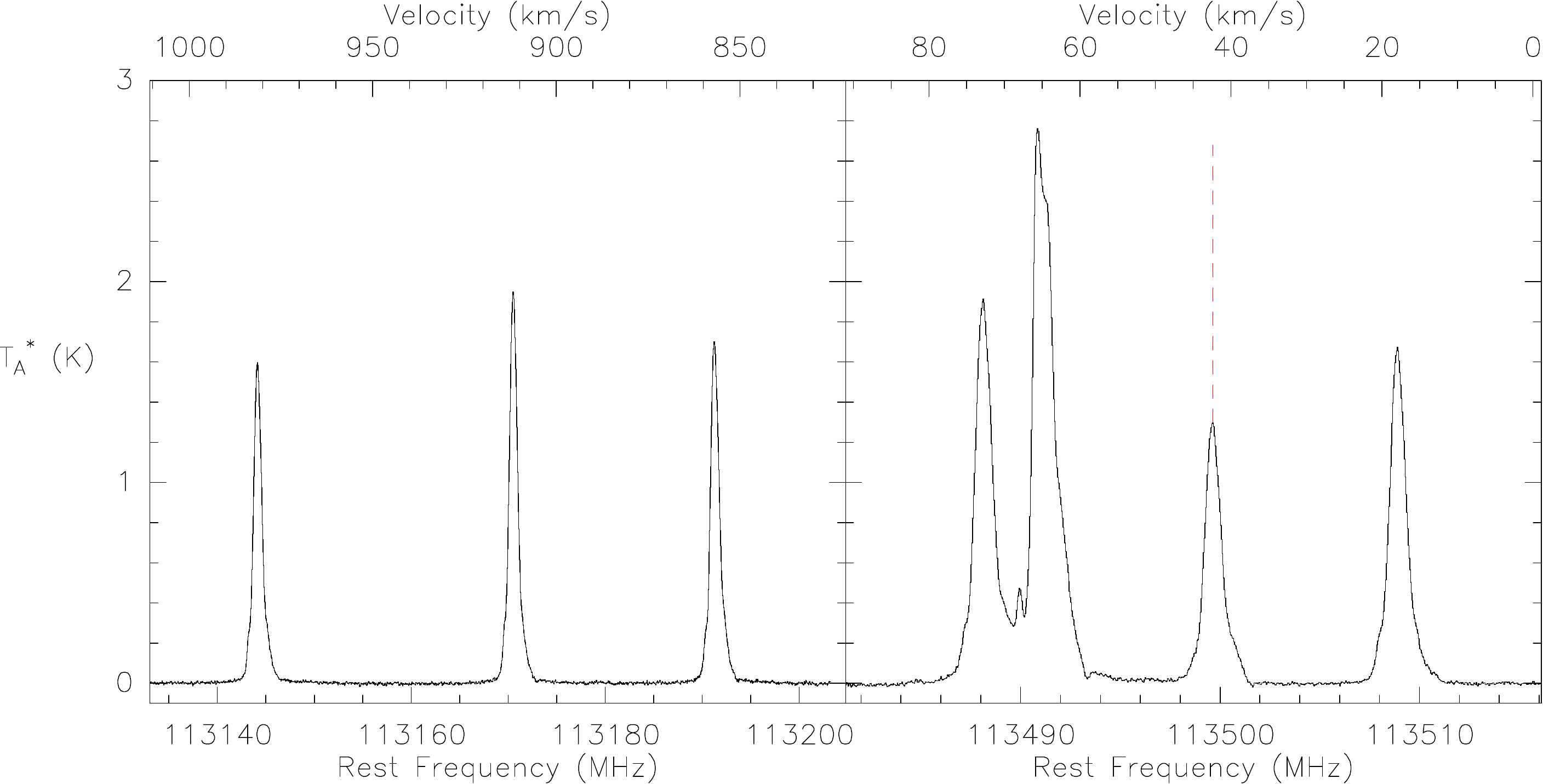}
\caption{CN(N $= 1 \rightarrow 0, {\rm J} = 1/2 \rightarrow 1/2$) (left)
  and ${\rm J} = 1/2 \rightarrow 3/2$ (right)  spectrum toward
  G35.20w. The LSR velocity is shown with respect to the component at 113499.64
  MHz (red dashed line). \label{fig:spec}}
\end{center}
\end{figure*}

In Fig.\ref{fig:spec}, we show our CN spectrum toward G35.20w. A
single velocity component yields a poor fit to the hyperfine spectrum. The
brightest component at 113488.12 GHz (line 4) is partially blended with the
component at 113490.97 GHz (line 5).  Excluding these components, we stack the
other hyperfine components and perform multiple component Gaussian fits (see
next section).

\begin{figure*}
\begin{center}
\begin{tabular}{cc}
\includegraphics[angle=0,scale=0.40]{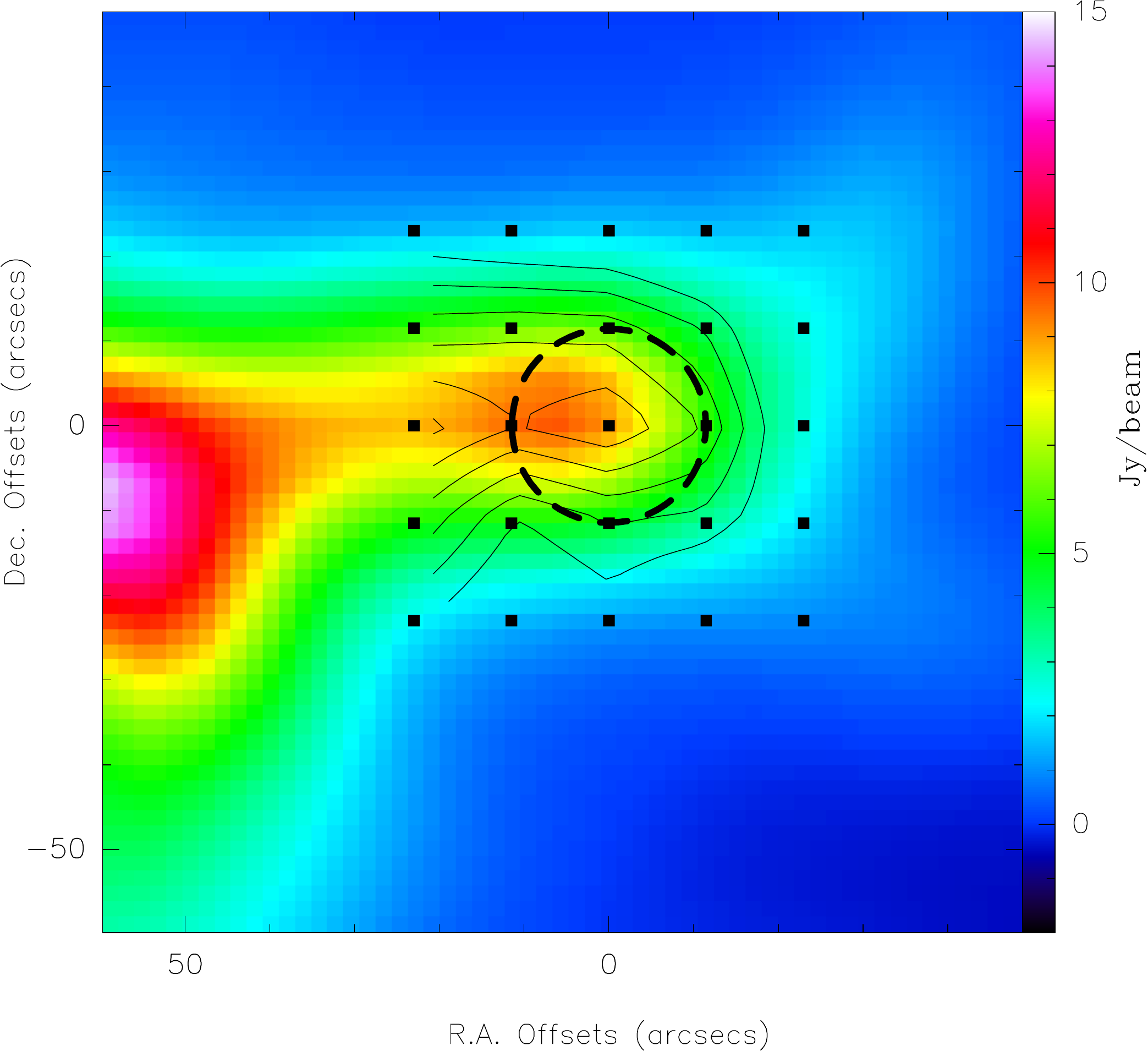} & 
\includegraphics[width=0.44\textwidth]{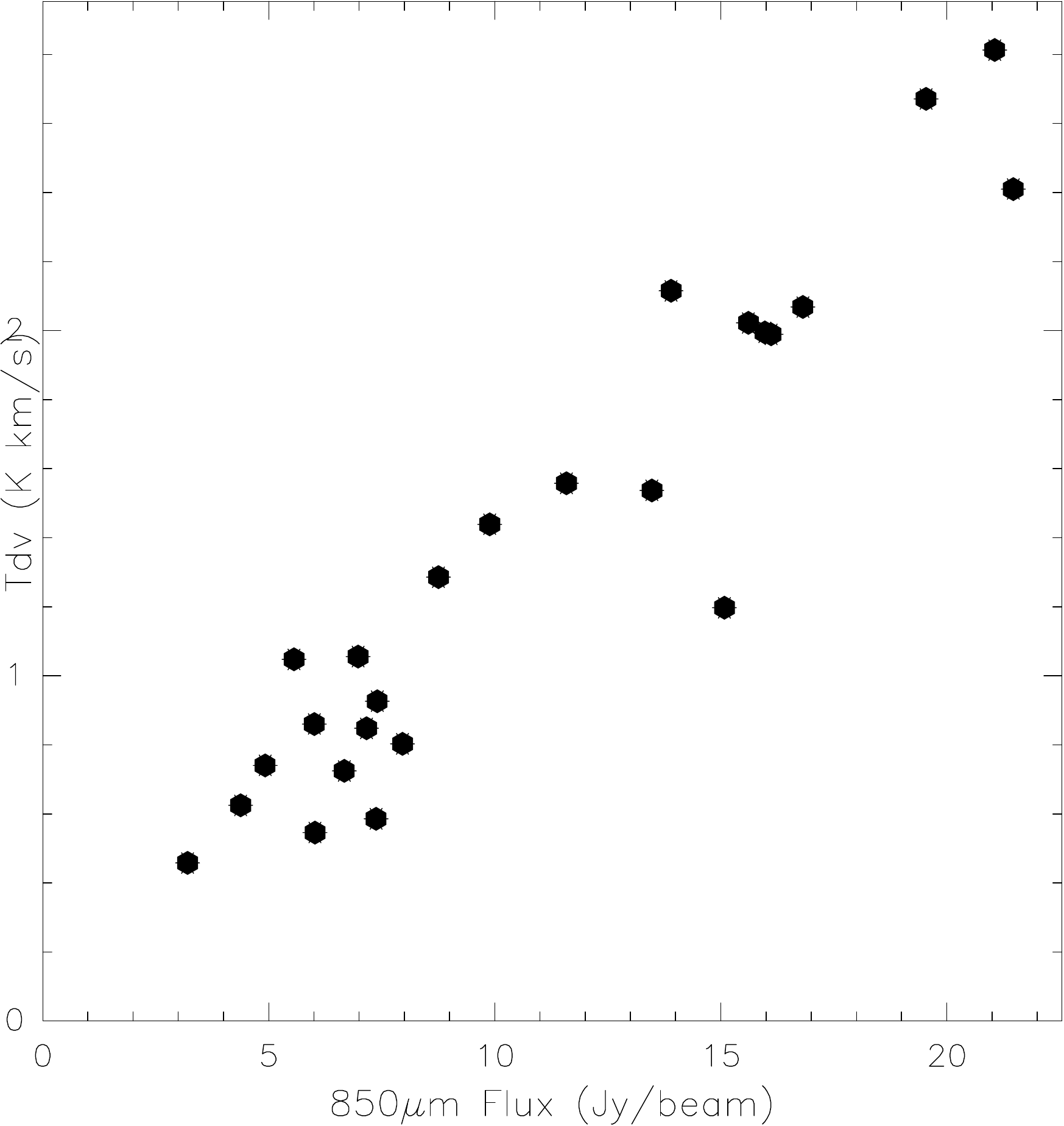}
\end{tabular}
\caption{Left: SCUBA  850\,$\mu$m dust continuum archival data (colorscale)
  and CN intensity integrated over the 113170.49 MHz hyperfine from 34
  to 42\,\kms (contours, 1.2 to 4 in steps of
  0.3~K\,\kms). The offsets positions for the CN map and the IRAM 30 m
  beam are shown as black squares and dashed circle
  respectively. Right: Correlation plot of 850\,$\mu$m flux (smoothed
  to 23.5\,\arcsec) and CN integrated intensity. \label{fig:cn_dust}}
\end{center}

\end{figure*}

Although the CN 1--0 transition has already been used to measure $B$-fields in active high-mass
regions (Crutcher et al. 2010), it has been unclear whether this transition is a
selective photodissociation region (PDR) tracer or rather a dense gas tracer in star-forming regions. A
recent wide-field interferometer mapping in the dense molecular
envelope of the ultracompact HII region W3OH, however, finds that CN is
an excellent tracer of warm high density gas \citep{hakobian2011}. 
\citet{maury2012:cn}  show that in NGC2264-C, an intermediate-mass
protocluster without significant ionizing radiation in its vicinity, the
CN emission closely follows the dust continuum emission. In
Fig.\,\ref{fig:cn_dust}, an integrated intensity map of CN for the
main hyperfine component is shown with respect to the SCUBA
850\,$\mu$m dust continuum emission. The dust continuum emission
within the field of view (FOV) shows a bright core at x,y-offset
$+60'',-10''$ (the well-known W48 UC\HII\ region) as well as a secondary core [at (0,0)] that is our target (G35.20w).  The CN emission
shows a compact peak at the 30 m pointing center and a structure that
correlates well with the dust emission as shown in  the correlation
plot in Fig.\,\ref{fig:cn_dust} (right panel).

In Figure~\ref{fig:zeeman-stack}, we show the final Stokes $V$
spectrum obtained as a weighted average of all Zeeman--sensitive
transitions. In spite of the deep observations with very good 3 mm
conditions, we do not detect a significant Stokes $V$ component. The
final rms is 4.6\,mK for a 0.1\,\kms\ velocity resolution for a
telescope  time of $\sim 33$ hours.

\subsubsection{CN 1-0 hyperfine fitting and optically depth  \label{sec:hfs_fit}}
In the following subsection, we discuss a new method to fit
the CN hfs components robustly and simultaneously. We stack all the
hyperfine components
except the blended lines (4 and 5 in Table~\ref{tab:freqs}) and obtain
a best fit with two velocity components.  A two-component fit provides
a good fit to the hfs components  with one narrow component ($\sim
2.1$\, \kms\ FWHM), with a peak brightness of 1.18~K centered at the systemic source velocity
(42.5\,\kms, see P11), and a weaker (0.46 K), wide ($\sim 5.0$\, \kms\
FWHM) and likely
diffuse component at $\sim 42.0$\,\kms.  We assume that it is the component
at the systemic velocity that is associated with the dense gas and
hence higher field strength that contributes to the Zeeman splitting. The
fit parameters (line width and velocity) are then saved into an
initialization file where the line width and velocity of the diffuse component is fixed. The input
file for our Zeeman analysis (next section \ref{sec:zeeman_fit}) has the
following parts.  The first part is obtained after we stack all the
hfs components together and perform two velocity component fits to the
stacked spectrum. We subtract both fits in the Stokes $I$ data based on the
initialization file.  The second part is obtained similarly, except
that we only subtract the diffuse feature at $\sim 42.0$\,\kms, where the
residual Stokes $I$ spectra is left with the systemic velocity feature
( $\sim 42.5$\,\kms). Finally, the original Stokes $I$ and Stokes $V$
are included as the third and fourth part.

Our approach to the Zeeman analysis relies on the assumption that the
emission of all the hfs components is optically thin. Previous CN 1-0 observations in even
more evolved regions confirm that the emission is optically thin
\citep{falgarone2008}.  The deviation of the relative intensity of the
hfs components from local thermodynamic equilibrium (LTE) is a measure of the optical depth. Following Table~1 of
\citet{falgarone2008}, for optically thick emission, the ratio of line
intensities for lines 6 and 7 with respect to line 5 is expected to be 1, while for small optical
depths, the expected ratio is 3.4.  Since the observed ratio of 2.1 is
intermediate to these values,
we conclude that the optical depth is moderate.  From the simultaneous
hfs component fit in CLASS, we also find that the optical depth for the brightest
hfs components (which we exclude from our analysis) is at most 1.4, while the weaker components have $\tau \ll
1$. Therefore, we proceed with the assumption that the emission is in LTE.

\subsubsection{CN Zeeman fitting  \label{sec:zeeman_fit}}

We follow \citet{crutcher1996} to fit the Zeeman signal in our
spectra. The velocity-dependent Stokes $I(v)$ and $V(v)$ signals of
the 1--0 transition of CN can be understood as a sum of seven hfs components. These are indicated as $I_i$ and $V_i$ in the
following. \citeauthor{crutcher1996} assume that $V_i$ can be
modeled as
\begin{equation}
\label{eq:Z-model}
V_i(v) = C_1 \cdot{} I_i(v) + C_2 \cdot{}
\frac{{\rm{}d}I_i(\nu)}{{\rm{}d}\,v} +
Z_i \cdot \frac{B_{\rm{}los}}{2} \cdot{}
\frac{{\rm{}d}\hat{I}_i(\nu)}{{\rm{}d}\,v} \, .
\end{equation}
The factors ${\rm{}d}I_i(\nu)/{\rm{}d}\,v$ and
${\rm{}d}\hat{I}_i(\nu)/{\rm{}d}\,v$ in this relation 
give the first
derivative of the spectra with respect to frequency. The Zeeman
splitting factors are given by the $Z_i$. We obtain $\hat{I}_i$ by
subtracting unrelated velocity components from the spectra $I_i$, as
described above. 

The first term in  Eq.~(\ref{eq:Z-model}) describes the instrumental,
quasi-achromatic leakage of Stokes I into Stokes V. The second term
accounts for artificial Zeeman features that arise when a velocity
gradient is seen by an anti-symmetric beam pattern in Stokes V, for example,
owing to a beam squint. \citet{thum2008} provide some typical beam shapes for different Stokes parameters.
The last term is proportional to the
magnetic flux density projected along the line of sight, $B_{\rm{}los}$. It describes the signal caused by the
Zeeman effect. A simultaneous fit of several hfs components in Stokes
$I_i$ and $V_i$ therefore, allows us to separate the true Zeeman signal in
the spectra from instrumental bias.For $Z_i$, we refer to Table~\ref{tab:freqs} 
(adapted from \citealt{falgarone2008}).  This method assumes
  that a velocity gradient if any is identical for both velocity
  components.  Given that the large scale component  is 
  much weaker and broader than the feature at the systemic velocity
  (see Section~\ref{sec:hfs_fit}), this is a valid assumption.

We distinguish between $I_i$ and $\hat{I}_i$ for the following
reasons. The CN spectra reveal two velocity
components. Only one of these components (42.5\,\kms) resides at the
systemic velocity of the dense gas (P11). It is plausible to assume
that only this velocity component, which is assumed to be described by
$\hat{I}_i$, produces a signal due to the Zeeman
effect. Equation~(\ref{eq:Z-model}) captures exactly this assumption
in the term proportional to $B_{\rm{}los}$. Signals from the other
velocity components can, however, leak into the system and cause
spurious emission. The terms proportional to $C_1$ and $C_2$, therefore,
include intensities $I_i$. The intensity $\hat{I}_i$ is found by
subtracting Gaussian components, which are centered at velocities offset from the
systemic one, from the observed intensities, $I_i$ as explained in
Section~\ref{sec:hfs_fit}. An inspection of $\hat{I}_i$ reveals that
the component subtraction is unsatisfying for the hfs component 5
because of blending. We therefore exclude this component from the
remainder of the analysis and work with a total of six hyperfine
satellites.

We use the MFIT routine within the GILDAS data processing package to
minimize the difference between the observed Stokes $V$ spectra and
the model in Eq.~(\ref{eq:Z-model}). In a first test of this approach
we fit Eq.~(\ref{eq:Z-model}) to spectra containing Gaussian noise
with a standard deviation of $4.6~\rm{}mK$, which is similar to that of
our observed spectrum, onto which artificial Zeeman signals following
Eq.~(\ref{eq:Z-model}) are superimposed. We vary the magnetic flux
density in the range $|B_{\rm{}los}|\le{}1~\rm{}mG$ for these
experiments and generate signals assuming $C_1=C_2=0$. For every value
of $B$ we produce and fit 2,000 spectra with different random
initializations. The magnetic flux densities derived from fitting
Eq.~(\ref{eq:Z-model}) to the synthetic spectra are then compared to
the values of $B_{\rm{}los}$ used to generate the spectra. We use
these experiments to explore a variety of weighting schemes, such as
weighting of velocity channels by the expected Zeeman signal,
$(Z_i\cdot{}{\rm{}d}\hat{I}_i[\nu]/{\rm{}d}\,v)^2$. This shows that,
among the options considered, the standard deviation between the input and
output value of $B_{\rm{}los}$ is minimized if the same weight is
given to all channels where $\hat{I}_i>0.2~\rm{}K$. Analysis of the
experiments shows that the magnetic field fitting results are centered
on the input value of $B_{\rm{}los}$, and that 68\% of the derived
values fall into the range $B_{\rm{}los}\pm{}420~\rm{}\mu{}G$. The
$1\sigma$ uncertainty in $B_{\rm{}los}$ is accordingly set to
$\sigma_{B_{\rm{}los}}=420~\rm{}\mu{}G$.

One interesting observation is the influence of the hfs component 2,
which is the only component with a negative Zeeman splitting
coefficient. Experiments with artificial spectra show that the
uncertainty increases to $\sigma({B_{\rm{}los}})=510$\muG\ if the sign
of $Z_{2}$ is flipped. We attribute this increase in the uncertainty
to the fact that the sign of $Z_{2}$ helps to separate spurious
components in $V_i$ from actual Zeeman signals. This can be gleaned
from Eq.~(\ref{eq:Z-model}); at a given velocity, $v$, spurious signals
due to ${\rm{}d}I_i/{\rm{}d}v$ bias $V_i$ in the same direction. This
is different for the actual Zeeman signal, where hfs component 2
causes a signal that is opposite in sign from those of all other
Zeeman-sensitive components.

\begin{figure}[h]
\includegraphics[angle=0,scale=0.32]{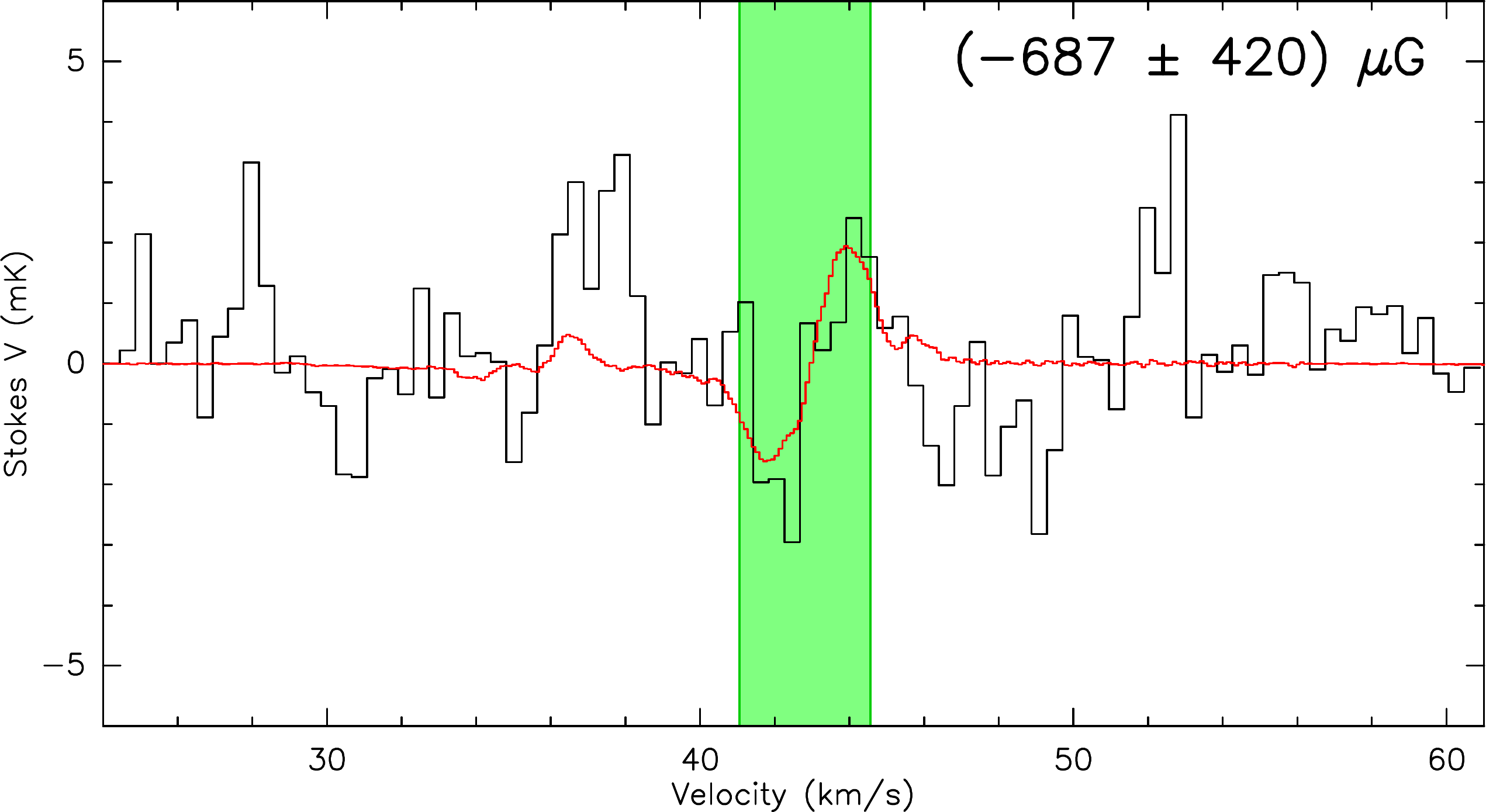}
\caption{Weighted average of all Zeeman-sensitive transitions
  (\emph{black spectrum}) after smoothing to $0.4~\rm{}km\,s^{-1}$
  velocity resolution, and best-fitting Zeeman signal (\emph{red
    spectrum}). The \emph{green highlighting} roughly indicates the
  spectral channels that we considered in the fitting
  process.\label{fig:zeeman-stack}}
\end{figure}

We then return to the actual observed $V$ spectrum and derive
$B_{\rm{}los}$ by fitting Eq.~(\ref{eq:Z-model}) to the data. This
yields
\begin{displaymath}
B_{\rm{}los} = (-687 \pm 420) ~ \rm{}\mu{}G \, ,
\end{displaymath}
where the uncertainty is taken from the aforementioned
experiments. This result is shown in Fig.~\ref{fig:zeeman-stack}. Here
we take the observed $V_i$ of all hyperfine satellites used in the
fit, we subtract the modeled spurious signals, we align all spectra in
velocity space, and we finally add them up to produce a stacked
observed spectrum. On this we overlay the model stacked Zeeman signal
summed up over all used hyperfine satellites. Both spectra are
slightly smoothed to a resolution of $0.4~\rm{}km\,s^{-1}$ to suppress
noise. We also roughly indicate the range of velocity channels used
for the analysis. This indeed suggests a possible Zeeman signal near
the perception threshold.\medskip

\noindent{}We briefly remark that our experiments with synthetic
spectra can be used to derive a general uncertainty estimate for
Zeeman observations using the CN 1-0 transition. For this, we assume
that ${\rm{}d}\hat{I}_i(\nu)/{\rm{}d}\,v$ depends on the line width
and the maximum intensity observed in the CN 1-0 transition as
${\rm{}d}\hat{I}_i(\nu)/{\rm{}d}\,v\propto{}\max(\hat{I})/\Delta{}v$. We
further assume that the uncertainty in $B_{\rm{}los}$ depends on the
noise in the spectra, $\sigma(V)$, which we assume to be similar to
$\sigma[I]$ and the first derivative of $I$ as
$\sigma(B_{\rm{}los})\propto{}\sigma(V)/({\rm{}d}\hat{I}_i(\nu)/{\rm{}d}\,v)$.
The number of useable channels (i.e., independent samples),
$N_{\rm{}chan}$, depends on the velocity resolution, $\delta{}v$, as
$N_{\rm{}chan}\propto{}\Delta{}v/\delta{}v$. Intuition suggests, and
experimentation confirms, that
$\sigma{}(B)\propto{}N_{\rm{}chan}^{-1/2}$.  In our case, the hyperfine
fitting discussed in section~\ref{sec:hfs_fit} yields the peak strength of
the stacked hyperfine components $\max(\hat{I})=1.18~\rm{}K$ and
$\Delta{}v=2.1~\rm{}km\,s^{-1}$ for the line component at systemic
velocity. We further have $\delta{}v=0.1~\rm{}km\,s^{-1}$,
$\sigma(V)=5~\rm{}mK$, and $\sigma(B_{\rm{}los})=420~\rm{}\mu{}G$ for
our observations. From this, we obtain
\begin{equation}
\begin{array}{rl}
\sigma(B_{\rm{}los}) & \displaystyle \approx{} 684~{\rm{}\mu{}G} \cdot{}
\left( \frac{\max(\hat{I})/\sigma(I)}{100} \right)^{-1} \\
 & \displaystyle \quad\quad\quad \cdot{} \left( \frac{\Delta{}v}{1~\rm{}km\,s^{-1}} \right)^{1/2} \cdot{}
\left( \frac{\delta{}v}{0.1~\rm{}km\,s^{-1}} \right)^{1/2} \, .
\end{array}
\end{equation}

\subsection{Dust polarization results}
The images presented in Figure~\ref{fig:polmap} summarize the results of the
archival dust polarization observations. The dust continuum emission is
shown in color scale with the plane of the sky (pos) component of the $B$-field (\Bpos)
overlaid. The length of the \Bpos\ vector represents the percentage
polarization, while we obtained its direction as shown by rotating the
polarization vector by $90{\degr}$. The $B$-field vectors are aligned perpendicular to the long filament connecting
our clump G35.20w with the W48 \HII\ region. High percentage
polarization ($\sim 6\%$) with remarkable alignment of the vectors
with the orientation of the  filament is seen over the whole extent of  G35.20w. There is neither evidence for a field distortion toward
the center of our clump nor a decrease in polarization
percentage. This is in stark contrast to the W48 \HII\ region itself,
which is located at an offset of $\sim 60$\,\arcsec, where the
dispersion in polarization angle is higher and polarization percentage
has dropped to $<1\%$ presumably because of the depolarization due
to the star formation process itself.

Employing the Chandrasekhar-Fermi (CF) method, the polarization
measurements allow us to estimate the projected field strength in the
plane of the sky.  The method relies on the assumption that the
magnetic, turbulent, and thermal pressure are in equipartition. Then
the plane-of-sky component, ${B}_{\rm{}pos}$, is
\citep{chandrasekhar1953}
\begin{equation}
 \label{eq:cf}
B_{\rm{}pos}=f\,\sqrt{4\pi\varrho}\,\frac{\sigma_v}{\sigma_{\phi}}\,.
\end{equation} Here, $\sigma_{\phi}$ is the dispersion in polarization
angle measured in radians and $\varrho$ is the mass density of the
region in the cloud relevant to the $\sigma_{\phi}$ and $\sigma_{v}$
values. A corrector factor $f=0.5$ is adopted from studies using
synthetic polarization maps generated from numerically simulated
clouds \citep{ostriker2001,heitsch01} as long as
$\sigma_{\phi}\le{}25^{\circ}$.

More sophisticated approaches have been discussed in the recent
years \citep{falceta-Goncalves2008,hildebrand2009,houde2009}. However, such methods require a statistically significant
number of measurements , a requirement that our data fail to satisfy. Moreover, the main
advantage of such treatments over the classical CF method is in
gauging the influence of the ordered large scale field due to
non-turbulent physics such as shock compression on the regular
field structure. This is clearly not an issue in G35.20w. 
We then use the density, velocity dispersion (Table~\ref{tab:mass}, see
section~\ref{sec:crit} for details), and angle dispersion
$\sigma_{\phi}=9.74 \ll 25^{\circ}$  (see Table\,\ref{tab:mag})
extracted for the box in Fig.1, and determine a pos field strength of
$742\pm{}247$\,\muG. This is consistent with the previous estimate of
$650$\,\muG\ by \citet{curran2004}.

 Table\,\ref{tab:mag}  contains statistical uncertainty estimates obtained via
  regular Gaussian error propagation. In addition, we also consider the
  impact of systematic uncertainties on relevant parameters. In our analysis we assume that
  $\sigma_v$ is dominated only by statistical uncertainties,
    while we assume that
  mass, $N_{\rm{}H_{2}}$ and $\varrho$ is uncertain by a
  factor 2. \citet{pillai2015} provide a detailed discussion of the error analysis.

\begin{table*}
\centering
\caption{Physical parameters \label{tab:mass}}
\begin{tabular}{ccccccccc}
\hline 
Source & Distance  &  S$_{850\mu m}$  & Radius & $\sigma_{v}$  & Mass &  N$_{H_{2}}$  & Density  & $\alpha$
 \tabularnewline
          & pc & Jy & pc & \kms\ & \Msol\ & (10$^{23}$\,\cmsq) & 10$^{5}$\,\percc &  \tabularnewline
\hline 
G35.20w   & 3270 &  33.6 & 0.35 & 0.93 & 1650 & 2.3 & 1.7 & 0.2  \tabularnewline
\hline 
\end{tabular} \\
Notes: Radius corresponds to that of an effective area  and $\sigma_{v}$ is  the gas velocity dispersion within the bounded box shown in Figure~\ref{fig:polmap} . 
\end{table*}

\begin{table*}
\centering
\caption{Polarization \& CN magnetic field parameters \label{tab:mag}}
\begin{tabular}{cccccccc}
\hline 
Source  &  $\sigma_{\phi}$ & \Bpos  & \Blos   & $\hbox{B}_{\rm tot}$  &  $\mathcal{M}_{\rm{}A}$ & $\frac{(M/\Phi_B)}{(M/\Phi_B)_{\rm{}cr}}$\\ 
            &rad & \muG\ & \muG\ &   $\mu$G & \tabularnewline
\hline 
G35.20w  &  $0.17\pm{}0.06$  & $742\pm{}247 \rvert_{452}^{1217}$ &
                                                                   $-687\pm{}420$
                         &  $1011 \rvert_{506}^{2022}$& $0.4
\rvert_{0.3}^{0.7}$ & $1.5\rvert_{0.8}^{2.7}$  \tabularnewline
\hline 
\end{tabular} \\
Notes: $\sigma_{\theta}$  is the dispersion in the polarization vector
within the bounded box shown in Figure~\ref{fig:polmap},  pos, los, and
total magnetic field ($\hbox{B}_{\rm pos}$, $\hbox{B}_{\rm los}$,$\hbox{B}_{\rm tot}$), Alfv\'en mach number
($\mathcal{M}_{\rm{}A}$), mass-to-flux parameter
$({M/\Phi_B})/{(M/\Phi_B)_{\rm{}cr}}$. A $\pm$ sign is adopted to show the statistical
uncertainty. Systematic uncertainties are given by the $1 \sigma$
estimate represented by the lower and upper bound, 
respectively, for the relevant parameters.

\end{table*}

%\subsection{Equilibrium Field Strength}

\section{Discussion}

 In this section we interpret the observed data in
context. Section~\ref{sec:est-vs-obs} compares different theoretical
and observational estimates of the observed field strengths. This in
particular includes a comparison between fields in the plane of the
sky to those along the line of sight. This requires a statistical
argument that is prepared in Sec.~\ref{sec:projection}. We finally
compare the estimated magnetic energy densities to those of
self--gravity and turbulent gas motions in
Sec.~\ref{sec:alfvenic-supercritical}.

\subsection{Projection of magnetic field strengths\label{sec:projection}}
The observational techniques applied to magnetic fields discussed here
can only detect projected components of the total field. Zeeman observations probe the
field components along the line of sight. The linear polarization
observations reveal the field perpendicular to the line
of sight. To infer the total field strength, we need to know how these
projected components relate to the total field strength.

We define  $\vartheta$ as the angle between the 3D magnetic
field vector, $\vec{B}$, and the plane of the sky. We define
$\vartheta$ to be positive when the field has components pointing away
from the observer. Defining $B\equiv{}|\vec{B}|$ as the absolute
field strength, the plane-of-sky and line-of-sight components of
the field are
\begin{subequations}
\label{eq:field-projections}
\begin{align}
  B_{\rm{}pos} & = B\cdot{}\cos(\vartheta) \\
  B_{\rm{}los} & = B\cdot{}\sin(\vartheta) \, .
\end{align}
\end{subequations}
Not all of these projections occur with the same probability. Consider
a vector $\vec{n}$ of normalized length 1 that describes a sphere. We allow
$\vartheta$ again to be the angle with respect to the plane of the sky as
described above. The end points of all vectors $\vec{n}$ with the same
$\vartheta$ describe a line on the sphere of length
$2\pi{}\cos({\vartheta})$. Consider a small change $\d{}\vartheta$ in
$\vartheta$. The moving line on the sphere then covers an area
$\d{}\Omega=2\pi\cos(\vartheta)\d\vartheta$. This area can be used to
determine the probability at which a specific projection on the sky
occurs. Consider the case where $\vec{n}$ points almost exactly away
from the observer. This is a very rare situation because there is
just a single point on the sphere that exactly satisfies this
criterion. Now consider the area. In this case, $\vartheta$ is close to
$90^\circ$ and $2\pi{}\cos(\vartheta)$ is close to zero. Thus
$\d{}\Omega$ is small, too. Now consider the case where $\vec{n}$ is
almost at a right angle with the line of sight. This is a much more
common situation: on the sphere there is a long line of positions with
a similar $\vartheta$, and $2\pi{}\cos(\vartheta)$ and $\d{}\Omega$
are close to their maximum.

\begin{figure}
\begin{center}

\includegraphics[angle=0,scale=0.60]{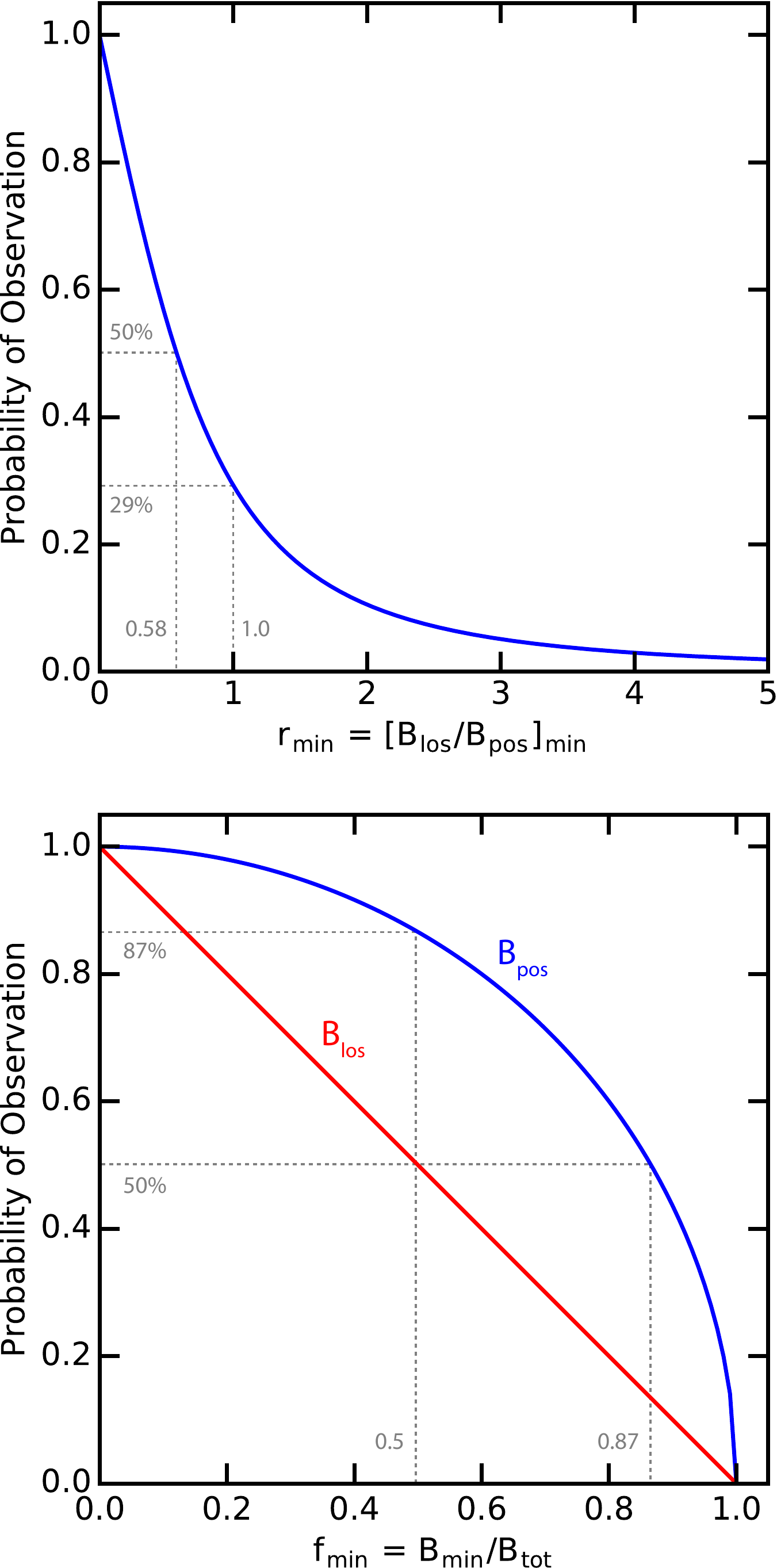} 
\caption{{\it Top panel:} probability for the ratio of the magnetic
  field measurement of the plane-of-sky component and of the
  line-of-sight component to yield a value of at least $r_{\rm min}$. {\it Bottom panel:} probability for a magnetic field measurement of the plane-of-sky component (blue curve)
and of the line-of-sight component (red curve) to yield a fraction of at least $f_{\rm min}$ of the
total magnetic flux density.\label{fig:projection-prob}}
\end{center}

\end{figure}

Here we are concerned with the absolute values of the projected
fields. In this situation, we can ignore the sign of $\vartheta$ and we
can limit the analysis to half of the sphere. We encode the probability
distribution of $\vartheta$ via the weighting function
$W(\vartheta)=2\pi{}\cos(\vartheta)$. The probability that projections with
$\vartheta$ are less than or equal to some angle $\vartheta_p$, $0\le \vartheta\le \vartheta_p$, is then
\begin{equation}
\label{eq:projection-probabiliy-ansatz}
p \equiv P(\vartheta\le{}\vartheta_p) =
  \frac{
    \int_0^{\vartheta_p}W(\vartheta)\d\vartheta
  }{
    \int_0^{90^\circ}W(\vartheta)\d\vartheta
  } \, .
\end{equation}
Evaluation yields
\begin{equation}
\label{eq:projection-probabiliy-result}
\vartheta_p = \arcsin(p) \, .
\end{equation}
We can use this result to evaluate the probability that projection
effects lead to a specific ratio between the projected field
strengths, $r=B_{\rm{}los}/B_{\rm{}pos}$. Substitution of
Eq.~(\ref{eq:field-projections}) gives $r=\tan(\vartheta)$. Here we
choose to display the probability that the actual field strength ratio
$r$ is larger than some minimum value $r_{\rm{}min}$,
$P(r>r_{\rm{}min})$. This essentially means that we use
$r_{\rm{}min}=\tan(\vartheta_p)$ to calculate the limiting angle
$\vartheta_p=\arctan(r_{\rm{}min})$ that is then used to obtain
$P(r>r_{\rm{}min})$ as $P(\vartheta\ge{}\vartheta_p)$, i.e., the
probability of having projection angles larger than $\vartheta_p$. By
definition, 
$P(\vartheta\ge{}\vartheta_p)=1-P(\vartheta\le{}\vartheta_p)=1-p$,
where we substituted Eq.~(\ref{eq:projection-probabiliy-ansatz}). We
then use $p=\sin(\vartheta_p)$ as a consequence of
Eq.~(\ref{eq:projection-probabiliy-result}) to eventually obtain
$P(r>r_{\rm{}min})=1-\sin(\arctan[r_{\rm{}min}])$.

For completeness, we extend this formalism also to estimate the
likelihood that a certain fraction of the magnetic field is projected
into the plane of the sky, respectively, along the line of
sight.  This aspect has already been discussed in detail by
\citet{heiles2005:lecture}. Assuming we wish to evaluate the probability that at least a
fraction $f_{\rm{}min}=B_{\rm{}pos}/B$ of the magnetic field is
projected into the plane of the
sky. Equation~(\ref{eq:field-projections}) gives the angle
$\vartheta=\arccos(f_{\rm{}min})$ that must not be exceeded to obtain
the required projected field
strength. Equation~(\ref{eq:projection-probabiliy-ansatz}) describes
the probability that this angle is not exceeded. Substitution then
gives the probability that $B_{\rm{}pos}$ exceeds a certain fraction
of the total field, $P_{\rm{}pos}=\sin(\arccos[f_{\rm{}min}])$. Now
assume we wish to determine the probability that at least a fraction
$f_{\rm{}min}=B_{\rm{}los}/B$ of the magnetic field is projected along
the line of sight. Now Eq.~(\ref{eq:field-projections}) gives the
angle $\vartheta=\arcsin(f_{\rm{}min})$ that must not \emph{fall short
  of} to obtain the required value of $B_{\rm{}los}$. Thus, we need to
know $P(\vartheta\ge{}\vartheta_p)$, which is by definition equal to
$1-P(\vartheta\le{}\vartheta_p)$. Substitution yields
$P_{\rm{}los}=1-\sin(\arcsin[f_{\rm{}min}])=1-f_{\rm{}min}$ as the
probability that $B_{\rm{}los}$ exceeds a fraction $f_{\rm{}min}$ of
the total field.

The probabilities $P_{\rm {}pos}(f_{\rm{}min})$ and
$P_{\rm{}los}(f_{\rm{}min})$ are shown in
Fig.~\ref{fig:projection-prob}. \emph{Note that $P_{\rm pos}$ always
  exceeds $P_{\rm{}los}$ by a significant factor.} In particular, in a
given situation it is always likely that $B_{\rm{}pos}$ represents a
significant fraction of the total field strength. For example, we see
that in 90\% of all cases $B_{\rm{}pos}$ is at least 44\% or more
of the true field strength $B$. $B_{\rm{}los}$, however, is more
likely to be significantly smaller than $B$;only in 50\% of all
situations is $B_{\rm{}los}$ at least 50\% of $B$.  This
discrepancy between the line-of-sight and plane-of-sky projections
is also seen in the mean projected field strengths. One finds
\begin{alignat}{2}
\langle{}B_{\rm{}pos}\rangle & =
  \frac{
    \int_0^{90^\circ}B_{\rm{}pos}(\vartheta)\,W(\vartheta)\d\vartheta
  }{
    \int_0^{90^\circ}W(\vartheta)\d\vartheta
  } = \frac{\pi}{4} B = 0.79\, B \label{eq:bpos}\\
\langle{}B_{\rm{}los}\rangle & =
  \frac{
    \int_0^{90^\circ}B_{\rm{}los}(\vartheta)\,W(\vartheta)\d\vartheta
  }{
    \int_0^{90^\circ}W(\vartheta)\d\vartheta
  } = \frac{1}{2} B \, . \label{eq:blos}
\end{alignat}
In general, \emph{plane-of-sky projections are much closer to the
  true field strength than line-of-sight projections}.

We note that Eq.~(\ref{eq:blos}) for \Blos\ is consistent with that derived
by \citet{heiles2005}, who showed that
statistically the average value of \Blos\ is only half of the total
field strength $B$.  However, Eq.~(\ref{eq:bpos})  shows that the same
correction factor cannot be applied for \Bpos\ (see also \citealt{heiles2005:lecture}).
Our result might explain the studies that so far have found
systematically higher field strengths based on Chandrasekhar-Fermi
method relative to Zeeman measurements
(e.g., \citealt{vallee2006,vallee2007a:omc1,vallee2007b:cb68,crutcher2012}.)

\subsection{Estimated vs.\ observed fields\label{sec:est-vs-obs}}
We are in the convenient situation that we  obtained two
perpendicular components of the magnetic fields, i.e., the
plane-of-sky and line-of-sight components. These independent
observations can be used to execute plausibility checks of these
observations. First, we study the ratio between \Blos\ and
\Bpos. This builds on the formalism derived in
Sec.~\ref{sec:projection}. We then compare the line-of-sight
component of the field to that obtained under the assumption that
the cloud core is balanced against self-gravity.

\subsubsection{Ratio between magnetic field components}
The formalism in Sec.~\ref{sec:projection} permits us to estimate how
likely certain ratios between field components are. Consider that we
obtain a plane-of-sky magnetic field strength \Bpos$=742$\muG\ and a
line-of-sight component of \Blos$=687$\,\muG. The ratio
$r=B_{\rm{}los}/B_{\rm{}pos}$ discussed above is thus $\sim 0.9$.
From Fig.~\ref{fig:projection-prob}~(top), we find that the probability of
detecting the orthogonal components of at least this value, $r\ge{}0.9$,
is 33\%. This is a very reasonable result.

We thus conclude that observations finding
$B_{\rm{}los}\approx{}B_{\rm{}pos}$, as obtained in our case,
are reasonable results. Values of $r$ very different from unity would
be unlikely to result from mere projection effects. In those
situations, it would be more likely that observational artifacts drive
the estimated field strengths.

We remark that dense cores with very high observed values of
plane-of-sky magnetic fields are not ideal candidates for Zeeman
detection experiments. As evident from
Fig.~\ref{fig:projection-prob}~(top), only in 29\% of all cases does one
expect that $B_{\rm{}los}$ exceeds $B_{\rm{}pos}$.
 Zeeman observations should be setup to be sensitive enough to
detect fields as faint as $0.58\cdot{}B_{\rm{}pos}$ to be able to
detect Zeeman signals with a probability $\ge 50$\%.

\subsubsection{Fields assuming balance against self--gravity}

We can also estimate the total magnetic field strength in the clump by
following a virial analysis.  For the clump properties tabulated in
Table~\ref{tab:mass}, we first gauge the turbulent support for the
cloud.  Section~\ref{sec:hfs_fit} shows that the CN emission
traces the dense gas and that it is optically thin. Therefore, we use
the CN velocity dispersion associated with the LSR velocity to measure
the kinetic energy.  We note that $\sigma_v=0.93$\,\kms\ is the sum of
the thermal $\rm{}H_2$ and the non--thermal motions of the bulk gas
(see Table~\ref{tab:mass}).
Then we follow Eq.~(1b) in \citet{kauffmann2013b} and find the virial
ratio $\alpha$ for turbulent support to be
$\alpha=0.2$. As discussed in detail in Kauffmann et al., this is
significantly lower than the critical value of $\sim 2$ for
stability. This suggests that the clump is likely to be highly
magnetized if stable. An ordered field perpendicular to the main axis is
strongly in favor of this scenario. We use Eq.~(16) of Kauffmann et
al. to estimate the field strength for a magnetized clump,
\begin{equation}
B = 81~{\rm{}\mu{}G} \, \frac{M_{\Phi}}{M_{\rm{}BE}} \,
  \left( \frac{\sigma_v}{\rm{}km\,s^{-1}} \right)^2
  \left( \frac{R}{\rm{}pc} \right)^{-1} \, .
\label{eq:magnetic-field-estimation}
\end{equation}
Following the definitions in Kauffmann et al., ${M_{\Phi}}$ is the magnetic flux mass and $M_{\rm{}BE}$ is the
Bonner-Ebert mass. Thus, we derive a total field strength of  1640\,\muG.  

Based on the results mentioned earlier, we infer that the line-of-sight magnetic
field is $687$\muG, while the plane-of-the-sky field shows a very ordered
structure perpendicular to the principal filament axis with an implied
field strength of $\sim 740$\muG.  Do the two complementary field estimates
contradict each other?
The expected line-of-sight  component based on virial analysis is
given by  \Blos /\muG\ $ =[1640^2-742^2]^{1/2}=1462$.
 \emph{This is in contradiction with our Zeeman measurement of 
$-687\pm420$\muG\ at a 1.8\,$\sigma$ level.}

Based on the probability distribution of orthogonal projections,
  we argued that the lack of a  Zeeman signal is not a
  surprising result. However, there are a few other reasons to question the
robustness of field determination using the methods above.  First, the CF method uses the dispersion in
the polarization angle. The observed polarization angle is an average
over several turbulent cells along the line of sight. The larger the
number of cells, the smoother the field and smaller the dispersion in
field angles. For our calculations, we assume a correction factor of 2
thatis consistent with sub-Alfv\'enic simulations, while
super-Alfv\'enic models suggest a larger correction.  However, super
Alfv\'enic turbulence would manifest itself as a chaotic and complex
field structure at odds with the observed smooth field
structure. Nevertheless, a correction factor of 3 or higher would
imply an even higher inconsistency with the Zeeman upper limit at
$>4\,\sigma$ level.  This rather suggests that field strengths derived
from polarization observations of this highly magnetized filament is
not an overestimate.  Second, systematic gas kinematics unrelated to
turbulence can align the field making it more ordered. No such large
scale field structure is evident in the polarization map toward
G35.20w.

As far as Zeeman measurements are concerned, first, there could be a
mutual cancellation of the Zeeman components within our 23.5\,\arcsec\
beam or exactly along our line-of-sight owing to field reversals. If the
latter is true, then even more sensitive observations like those with
ALMA will fail to detect any signal. Simulations by
\citet{bertram2012} show that even for sub-Alfv\'enic turbulence,
unfortunate angles between the line of sight and the mean field
direction can cause significant field reversals. Second, depletion of
CN at the very center could weaken the detection considerably
\citep{tassis2012b}.  P11 report \DAMM\ emission with high deuterium
fractionation that is indicative of cold conditions conducive to
depletion of heavy neutrals in G35.20w.  However, similar to the
results for low-mass prestellar cores \citep{blant2008}, we also do
not find a depletion hole in the CN integrated intensity map (see
Fig.\, 2). However, we cannot rule out an abundance drop on smaller
core scales.

\subsection{Sub-Alfv\'enic and super-critical\label{sec:alfvenic-supercritical}}
Polarization measurements of a range of dense ISM structures, from
pristine high-mass filaments \citep{pillai2015} to high-mass star-forming cores \citep{girart2009,zhang2014:bfields,li2015:bfield}, have
shown that magnetic fields play a dominant role from cloud to core
scales.

A direct measurement of the implied strong field however, has been
largely evasive. CN Zeeman observations in star-forming regions allow
for the most direct determination of field strength in dense gas
\citep{crutcher1996}. Robust CN Zeeman observations to date have been
made toward active high-mass star-forming regions as opposed to
younger evolutionary phases. We compared our Zeeman measurements with
a recent compilation of B-field estimates in more evolved high-mass
regions \citep{falgarone2008}. Using \Blos\ as an approximation to the
total field, Falgarone et al. find that protostellar clumps are
critical to moderately super-critical. The threshold value is defined
based on the mass-to-flux ratio $M/\Phi_B$, where the magnetic flux
$\Phi_B=\pi\,\langle{}B\rangle{}\,R^2$ is derived from the mean
magnetic field and the radius of the cloud cross-section
perpendicular to the magnetic field. The critical mass-to-flux ratio
is \citep{nakano1978} $(M/\Phi_B)_{\rm{}cr}=1/(2 \pi G^{1/2})$, where
$G$ is the gravitational constant.  We calculate this ratio in terms
of the observed column density and magnetic field following Eqn.(7) of
\citet{pillai2015}.  We use
$B_{\rm{}tot}=[B_{\rm{}pos}^2+B_{\rm{}los}^2]^{1/2}=1011$\,\muG\
as the total magnetic field.  A clump is magnetically sub-critical if
$(M/\Phi_B)<(M/\Phi_B)_{\rm{}cr}$.  We derive a ratio of 1.5 (see
Table~\ref{tab:mag}), which is consistent with the mean mass-to-flux
ratio derived by \citet{falgarone2008} for star-forming clumps, but
lower than the value derived by \citep{pillai2015} for pristine
high-mass regions. This is expected because magnetized support in the
early stages gives way to initial collapse as the clump evolves. Once
accretion onto protostars is initiated, gravity overwhelms magnetic
support, changing the field from sub-critical to super-critical.

What about turbulence? The Alfv\'en mach number is given by
$\mathcal{M}_{\rm{}A}=\sqrt{3}\,\sigma_{v}/v_{\rm{}A}$ , where
$v_{\rm{}A}=B_{\rm{}tot}/\sqrt{4\,\pi\,\varrho}$ is the Alfv\'en
speed. We find that even though
turbulence is supersonic in the clump, $\mathcal{M}_{\rm{}A}=0.4$
and, thus, sub-Alfv\'enic. This is similar to the results for the
starless high-mass regions in \citep{pillai2015}, while
\citet{falgarone2008}  find their more evolved cores to be in the
Alfv\'enic to super-Alfv\'enic domain. Localized feedback associated
with star formation especially in the active accretion and ejection
phase can lead to enhanced turbulence and, therefore, modify the
dynamics from magnetic field dominated to an equipartition state.  
This evolution manifests  in the form of ordered
field structure in the early stages that are increasingly perturbed by enhanced
turbulence and gravity to chaotic field orientations. 

\section{Summary}
G35.20w is a young high-mass clump in W48 that appears to be
undergoing collapse unless supported by a very strong field of
1640\muG\ (Section 4.1). Our CN $N=1-0$ Zeeman effect measurement
toward G35.20w yields a line-of-sight field strength of
$-687 \pm 420$\muG\ (Section 4.2.2).  Thermal dust continuum
polarization indicates that the magnetic field direction is aligned
with the minor axis of the clump.  The dispersion in polarization
angles also provides a plane-of-sky component of the magnetic field of
$\sim 740$\muG\ (Section 4.3). We show that plane-of-sky projections
are much closer to the true field strengths than line-of-sight
projections (Fig.~\ref{fig:projection-prob}). The mass-to-flux ratio
exceeds its critical value by a factor 1.5, indicating considerable
magnetic support. Forces induced by the magnetic field significantly
exceed those owing to the turbulent support ($\mathcal{M}_{\rm{}A}=0.4$).

%______________________________________________________________

\begin{acknowledgements}
       We thank the IRAM 30 m staff for hosting us. We thank Kazi Rygl for
        providing us Herschel based mass and temperature maps. T.P. acknowledges support
      from the \emph{Deut\-sche For\-schungs\-ge\-mein\-schaft, DFG\/}
      via the SPP (priority program) 1573 ‘Physics of the ISM’.   TP \&
      JK thank Nissim Kanekar for an invitation to the National Centre for Radio Astrophysics (NCRA) in
   Pune, India where some parts of this project were completed. We
   thank the anonymous referee for pointing out the reference to the
   work by Heiles and Crutcher 2005 during the revision of this manuscript.

\end{acknowledgements}

%% The following command ends your manuscript. LaTeX will ignore any text
%% that appears after it.
\bibliographystyle{aa}
%\bibliography{bib_astro.bib}

\begin{thebibliography}{42}
\expandafter\ifx\csname natexlab\endcsname\relax\def\natexlab#1{#1}\fi

\bibitem[{{Bertram} {et~al.}(2012){Bertram}, {Federrath}, {Banerjee}, \&
  {Klessen}}]{bertram2012}
{Bertram}, E., {Federrath}, C., {Banerjee}, R., \& {Klessen}, R.~S. 2012,
  \mnras, 420, 3163

\bibitem[{{Chandrasekhar} \& {Fermi}(1953)}]{chandrasekhar1953}
{Chandrasekhar}, S. \& {Fermi}, E. 1953, \apj, 118, 116

\bibitem[{{Crutcher}(1999)}]{crutcher1999}
{Crutcher}, R.~M. 1999, \apj, 520, 706

\bibitem[{{Crutcher}(2012)}]{crutcher2012}
{Crutcher}, R.~M. 2012, \araa, 50, 29

\bibitem[{{Crutcher} {et~al.}(1996){Crutcher}, {Troland}, {Lazareff}, \&
  {Kazes}}]{crutcher1996}
{Crutcher}, R.~M., {Troland}, T.~H., {Lazareff}, B., \& {Kazes}, I. 1996, \apj,
  456, 217

\bibitem[{{Curran} {et~al.}(2004){Curran}, {Chrysostomou}, {Collett},
  {Jenness}, \& {Aitken}}]{curran2004}
{Curran}, R.~L., {Chrysostomou}, A., {Collett}, J.~L., {Jenness}, T., \&
  {Aitken}, D.~K. 2004, \aap, 421, 195

\bibitem[{{Falceta-Gon{\c c}alves} {et~al.}(2008){Falceta-Gon{\c c}alves},
  {Lazarian}, \& {Kowal}}]{falceta-Goncalves2008}
{Falceta-Gon{\c c}alves}, D., {Lazarian}, A., \& {Kowal}, G. 2008, \apj, 679,
  537

\bibitem[{{Falgarone} {et~al.}(2008){Falgarone}, {Troland}, {Crutcher}, \&
  {Paubert}}]{falgarone2008}
{Falgarone}, E., {Troland}, T.~H., {Crutcher}, R.~M., \& {Paubert}, G. 2008,
  \aap, 487, 247

\bibitem[{{Girart} {et~al.}(2009){Girart}, {Beltr{\'a}n}, {Zhang}, {Rao}, \&
  {Estalella}}]{girart2009}
{Girart}, J.~M., {Beltr{\'a}n}, M.~T., {Zhang}, Q., {Rao}, R., \& {Estalella},
  R. 2009, Science, 324, 1408

\bibitem[{{Hakobian} \& {Crutcher}(2011)}]{hakobian2011}
{Hakobian}, N.~S. \& {Crutcher}, R.~M. 2011, \apj, 733, 6

\bibitem[{{Heiles} \& {Crutcher}(2005)}]{heiles2005:lecture}
{Heiles}, C. \& {Crutcher}, R. 2005, in Lecture Notes in Physics, Berlin
  Springer Verlag, Vol. 664, Cosmic Magnetic Fields, ed. R.~{Wielebinski} \&
  R.~{Beck}, 137

\bibitem[{{Heiles} \& {Troland}(2005)}]{heiles2005}
{Heiles}, C. \& {Troland}, T.~H. 2005, \apj, 624, 773

\bibitem[{{Heitsch} {et~al.}(2001){Heitsch}, {Zweibel}, {Mac Low}, {Li}, \&
  {Norman}}]{heitsch01}
{Heitsch}, F., {Zweibel}, E.~G., {Mac Low}, M.-M., {Li}, P., \& {Norman}, M.~L.
  2001, \apj, 561, 800

\bibitem[{{Hildebrand} {et~al.}(2009){Hildebrand}, {Kirby}, {Dotson}, {Houde},
  \& {Vaillancourt}}]{hildebrand2009}
{Hildebrand}, R.~H., {Kirby}, L., {Dotson}, J.~L., {Houde}, M., \&
  {Vaillancourt}, J.~E. 2009, \apj, 696, 567

\bibitem[{{Hily-Blant} {et~al.}(2008){Hily-Blant}, {Walmsley}, {Pineau Des
  For{\^e}ts}, \& {Flower}}]{blant2008}
{Hily-Blant}, P., {Walmsley}, M., {Pineau Des For{\^e}ts}, G., \& {Flower}, D.
  2008, \aap, 480, L5

\bibitem[{{Houde} {et~al.}(2009){Houde}, {Vaillancourt}, {Hildebrand},
  {Chitsazzadeh}, \& {Kirby}}]{houde2009}
{Houde}, M., {Vaillancourt}, J.~E., {Hildebrand}, R.~H., {Chitsazzadeh}, S., \&
  {Kirby}, L. 2009, \apj, 706, 1504

\bibitem[{{Kauffmann} {et~al.}(2008){Kauffmann}, {Bertoldi}, {Bourke}, {Evans},
  \& {Lee}}]{kauffmann2008}
{Kauffmann}, J., {Bertoldi}, F., {Bourke}, T.~L., {Evans}, II, N.~J., \& {Lee},
  C.~W. 2008, \aap, 487, 993

\bibitem[{{Kauffmann} \& {Pillai}(2010)}]{kauffmann2010c}
{Kauffmann}, J. \& {Pillai}, T. 2010, \apjl, 723, L7

\bibitem[{{Kauffmann} {et~al.}(2013){Kauffmann}, {Pillai}, \&
  {Goldsmith}}]{kauffmann2013b}
{Kauffmann}, J., {Pillai}, T., \& {Goldsmith}, P.~F. 2013, \apj, 779, 185

\bibitem[{{Kauffmann} {et~al.}(2010){Kauffmann}, {Pillai}, {Shetty}, {Myers},
  \& {Goodman}}]{kauffmann2010a:mass_size1}
{Kauffmann}, J., {Pillai}, T., {Shetty}, R., {Myers}, P.~C., \& {Goodman},
  A.~A. 2010, \apj, 712, 1137

\bibitem[{{Li} {et~al.}(2014){Li}, {Goodman}, {Sridharan}, {Houde}, {Li},
  {Novak}, \& {Tang}}]{li2014:ppvi_bfield}
{Li}, H.-B., {Goodman}, A., {Sridharan}, T.~K., {et~al.} 2014, Protostars and
  Planets VI, 101

\bibitem[{{Li} {et~al.}(2015){Li}, {Yuen}, {Otto}, {Leung}, {Sridharan},
  {Zhang}, {Liu}, {Tang}, \& {Qiu}}]{li2015:bfield}
{Li}, H.-B., {Yuen}, K.~H., {Otto}, F., {et~al.} 2015, \nat, 520, 518

\bibitem[{{Matthews} {et~al.}(2009){Matthews}, {McPhee}, {Fissel}, \&
  {Curran}}]{matthews2009}
{Matthews}, B.~C., {McPhee}, C.~A., {Fissel}, L.~M., \& {Curran}, R.~L. 2009,
  \apjs, 182, 143

\bibitem[{{Maury} {et~al.}(2012){Maury}, {Wiesemeyer}, \&
  {Thum}}]{maury2012:cn}
{Maury}, A.~J., {Wiesemeyer}, H., \& {Thum}, C. 2012, \aap, 544, A69

\bibitem[{{McKee} \& {Tan}(2002)}]{mckee02:100000yrs}
{McKee}, C.~F. \& {Tan}, J.~C. 2002, \nat, 416, 59

\bibitem[{{Nakano} \& {Nakamura}(1978)}]{nakano1978}
{Nakano}, T. \& {Nakamura}, T. 1978, \pasj, 30, 671

\bibitem[{{Ossenkopf} \& {Henning}(1994)}]{ossenkopf1994:opacities}
{Ossenkopf}, V. \& {Henning}, T. 1994, \aap, 291, 943

\bibitem[{{Ostriker} {et~al.}(2001){Ostriker}, {Stone}, \&
  {Gammie}}]{ostriker2001}
{Ostriker}, E.~C., {Stone}, J.~M., \& {Gammie}, C.~F. 2001, \apj, 546, 980

\bibitem[{{Pillai} {et~al.}(2015){Pillai}, {Kauffmann}, {Tan}, {Goldsmith},
  {Carey}, \& {Menten}}]{pillai2015}
{Pillai}, T., {Kauffmann}, J., {Tan}, J.~C., {et~al.} 2015, \apj, 799, 74

\bibitem[{{Pillai} {et~al.}(2011){Pillai}, {Kauffmann}, {Wyrowski}, {Hatchell},
  {Gibb}, \& {Thompson}}]{pillai2011a}
{Pillai}, T., {Kauffmann}, J., {Wyrowski}, F., {et~al.} 2011, \aap, 530, A118+

\bibitem[{{Rygl} {et~al.}(2014){Rygl}, {Goedhart}, {Polychroni}, {Wyrowski},
  {Motte}, {Elia}, {Nguyen-Luong}, {Didelon}, {Pestalozzi}, {Benedettini},
  {Molinari}, {Andr{\'e}}, {Fallscheer}, {Gibb}, {Giorgio}, {Hill},
  {K{\"o}nyves}, {Marston}, {Pezzuto}, {Rivera-Ingraham}, {Schisano},
  {Schneider}, {Spinoglio}, {Ward-Thompson}, \& {White}}]{rygl2014}
{Rygl}, K.~L.~J., {Goedhart}, S., {Polychroni}, D., {et~al.} 2014, \mnras, 440,
  427

\bibitem[{{Tan} {et~al.}(2014){Tan}, {Beltr{\'a}n}, {Caselli}, {Fontani},
  {Fuente}, {Krumholz}, {McKee}, \& {Stolte}}]{tan2014}
{Tan}, J.~C., {Beltr{\'a}n}, M.~T., {Caselli}, P., {et~al.} 2014, Protostars
  and Planets VI, 149

\bibitem[{{Tan} {et~al.}(2013){Tan}, {Kong}, {Butler}, {Caselli}, \&
  {Fontani}}]{tan2013:hmsc}
{Tan}, J.~C., {Kong}, S., {Butler}, M.~J., {Caselli}, P., \& {Fontani}, F.
  2013, \apj, 779, 96

\bibitem[{{Tassis} {et~al.}(2012){Tassis}, {Willacy}, {Yorke}, \&
  {Turner}}]{tassis2012b}
{Tassis}, K., {Willacy}, K., {Yorke}, H.~W., \& {Turner}, N.~J. 2012, \apj,
  754, 6

\bibitem[{{Thompson} {et~al.}(2005){Thompson}, {Gibb}, {Hatchell}, {Wyrowski},
  \& {Pillai}}]{thompson2005:dmuconf}
{Thompson}, M.~A., {Gibb}, A.~G., {Hatchell}, J.~H., {Wyrowski}, F., \&
  {Pillai}, T. 2005, in The Dusty and Molecular Universe: A Prelude to Herschel
  and ALMA, 425--426

\bibitem[{{Thum} {et~al.}(2008){Thum}, {Wiesemeyer}, {Paubert}, {Navarro}, \&
  {Morris}}]{thum2008}
{Thum}, C., {Wiesemeyer}, H., {Paubert}, G., {Navarro}, S., \& {Morris}, D.
  2008, \pasp, 120, 777

\bibitem[{{Vall{\'e}e} \& {Fiege}(2006)}]{vallee2006}
{Vall{\'e}e}, J.~P. \& {Fiege}, J.~D. 2006, \apj, 636, 332

\bibitem[{{Vall{\'e}e} \& {Fiege}(2007{\natexlab{a}})}]{vallee2007a:omc1}
{Vall{\'e}e}, J.~P. \& {Fiege}, J.~D. 2007{\natexlab{a}}, \aj, 133, 1012

\bibitem[{{Vall{\'e}e} \& {Fiege}(2007{\natexlab{b}})}]{vallee2007b:cb68}
{Vall{\'e}e}, J.~P. \& {Fiege}, J.~D. 2007{\natexlab{b}}, \aj, 134, 628

\bibitem[{{Zhang} {et~al.}(2009){Zhang}, {Zheng}, {Reid}, {Menten}, {Xu},
  {Moscadelli}, \& {Brunthaler}}]{zhang09_g35}
{Zhang}, B., {Zheng}, X.~W., {Reid}, M.~J., {et~al.} 2009, \apj, 693, 419

\bibitem[{{Zhang} {et~al.}(2014){Zhang}, {Qiu}, {Girart}, {(Baobab Liu},
  {Tang}, {Koch}, {Li}, {Keto}, {Ho}, {Rao}, {Lai}, {Ching}, {Frau}, {Chen},
  {Li}, {Padovani}, {Bontemps}, {Csengeri}, \&
  {Ju{\'a}rez}}]{zhang2014:bfields}
{Zhang}, Q., {Qiu}, K., {Girart}, J.~M., {et~al.} 2014, \apj, 792, 116

\bibitem[{{Zinnecker} \& {Yorke}(2007)}]{zinnecker2007:araa}
{Zinnecker}, H. \& {Yorke}, H.~W. 2007, \araa, 45, 481

\end{thebibliography}

\end{document}